\pdfoutput=1
\documentclass[prl,twocolumn,showpacs,nofootinbib]{revtex4-1}
\usepackage{graphicx,amsmath,amssymb,amscd,amsfonts,mathtools,moresize}
\usepackage[hidelinks]{hyperref}
\usepackage{xcolor}
\usepackage{ulem} 
\usepackage{thmtools, thm-restate}
\usepackage{epsfig}
\usepackage{epstopdf}
\usepackage{latexsym}
\usepackage{graphicx}
\usepackage{booktabs}
\usepackage{bbm}
\usepackage{color}
\usepackage{physics}
\usepackage{tensor}
\usepackage{verbatim}
\usepackage[caption=false]{subfig}
\usepackage{tikz}
\usepackage{bigints}
\usepackage{ifthen}
\usetikzlibrary{matrix}
\usetikzlibrary{decorations.markings,calc,shapes,decorations.pathmorphing,arrows.meta}
\usetikzlibrary{patterns}
\usetikzlibrary{positioning}
\usepackage{textpos}

\usepackage{hyperref}

\usepackage{xcolor}

\hypersetup{
colorlinks,
linkcolor={blue!80!black},
citecolor={blue!80!black},
urlcolor={blue!80!black},
linktoc=page
}

\usepackage{graphicx}
\usepackage{amssymb}
\usepackage{amsmath,amssymb}
\usepackage{slashed}
\usepackage{caption}
\usepackage{xcolor}
\usepackage{dsfont}
\usepackage{verbatim}
\usepackage{subfig}
\usepackage{amsmath,amssymb,amscd,amsfonts,mathtools}
\usepackage{xifthen}
\usepackage{xcolor}
\usepackage[toc,page]{appendix}
\usepackage{epsfig}
\usepackage{epstopdf}
\usepackage{latexsym}
\usepackage{graphicx}
\usepackage{booktabs}
\usepackage{bbm}
\usepackage{color}
\usepackage{physics}
\usepackage{tensor}
\usepackage{verbatim}
\usepackage{subfig}
\usepackage{tikz}
\usepackage{ifthen}
\usetikzlibrary{matrix}
\usetikzlibrary{decorations.markings,calc,shapes,decorations.pathmorphing,patterns,decorations.pathreplacing,arrows.meta}
\usetikzlibrary{positioning}
\usepackage{hyperref}

\usetikzlibrary{patterns}
\usetikzlibrary{positioning}

\pdfoutput=1
\makeatletter
\def\@fpheader{\relax}
\newcommand*{\ov}[1]{%
  $\m@th\overline{\mbox{#1}}$%
}
\newcommand*{\ovA}[1]{%
  $\m@th\overline{\mbox{#1}\raisebox{3mm}{}}$%
}
\newcommand*{\ovB}[1]{%
  $\m@th\overline{\mbox{#1\rule{0pt}{3mm}}}$%
}
\newcommand*{\ovC}[1]{%
  $\m@th\overline{\mbox{#1\strut}}$%
}
\newcommand*{\ovD}[1]{%
  $\m@th\overline{\mbox{#1\vphantom{\"A}}}$%
}
\newcommand*{\ovE}[1]{%
  $\m@th\overline{\raisebox{0pt}[1.2\height]{#1}}$%
}
\newcommand*{\ovF}[1]{%
  $\m@th\overline{\raisebox{0pt}[\dimexpr\height+1mm\relax]{#1}}$%
}
\newcommand*{\ovG}[1]{%
  $\m@th\overline{\raisebox{0pt}[\dimexpr\height+1mm\relax]{#1\vphantom{A}}}$%
}
\begin{document}
\def\nn{\nonumber}
\def\kc#1{\left(#1\right)}
\def\kd#1{\left[#1\right]}
\def\ke#1{\left\{#1\right\}}
\newcommand\beq{\begin{equation}}
\newcommand\eeq{\end{equation}}
\renewcommand{\Re}{\mathop{\mathrm{Re}}}
\renewcommand{\Im}{\mathop{\mathrm{Im}}}
\renewcommand{\b}[1]{\mathbf{#1}}
\renewcommand{\c}[1]{\mathcal{#1}}
\renewcommand{\u}{\uparrow}
\renewcommand{\d}{\downarrow}
\newcommand{\be}{\begin{equation}}
\newcommand{\ee}{\end{equation}}
\newcommand{\bsigma}{\boldsymbol{\sigma}}
\newcommand{\blambda}{\boldsymbol{\lambda}}
\newcommand{\sgn}{\mathop{\mathrm{sgn}}}
\newcommand{\diag}{\mathop{\mathrm{diag}}}
\newcommand{\Pf}{\mathop{\mathrm{Pf}}}
\newcommand{\half}{{\textstyle\frac{1}{2}}}
\newcommand{\sh}{{\textstyle{\frac{1}{2}}}}
\newcommand{\ish}{{\textstyle{\frac{i}{2}}}}
\newcommand{\thf}{{\textstyle{\frac{3}{2}}}}
\newcommand{\SUN}{SU(\mathcal{N})}
\newcommand{\N}{\mathcal{N}}

\renewcommand{\d}{\ensuremath{\operatorname{d}\!}}
\newcommand\dnote[1]{\textcolor{red}{\bf [Daniel:\,#1]}}

\title{\large Covariant Locally Localized Gravity and vDVZ Continuity}
\preprint{today}
\begin{abstract}
 {The Karch-Randall braneworld concerns the physics of an AdS$_{d}$ brane embedded in an ambient gravitational AdS$_{d+1}$ spacetime. The gravitational theory induced on the AdS$_{d}$ brane has a very light but massive graviton. It has been established that the zero graviton mass limit of the $d$-dimensional graviton propagator is smooth at tree-level. Furthermore, this smoothness was conjectured to persist to the quantum level. This conjecture suggests that the massive graviton on the AdS$_{d}$ brane is due to spontaneous symmetry breaking, which is consistent with its holographic dual description. In this letter, we show that the zero mass limit of the partition function is a theory of a massless graviton and a decoupled massive vector. The zero mass limit is not the basic Randall-Sundrum II model, but a theory with these additional decoupled vector degrees of freedom coupled only to gravity. The proof relies on deriving the fully covariant description of the $d$-dimensional gravity theory which enables us to compute the one-loop partition function. At the end, we comment on the implications of this result to the physics of entanglement islands.}
\end{abstract}

\author{Hao Geng$^{a}$, Moritz Merz$^{b}$ and Lisa Randall$^{a}$}

\affiliation{$^{a}$Gravity, Spacetime, and Particle Physics (GRASP) Initiative, Harvard University, 17 Oxford Street, Cambridge, MA 02138, USA.\\
$^{b}$Harvard Quantum Initiative, Harvard University, Cambridge, MA 02138 USA.}
\maketitle
\section{Introduction}
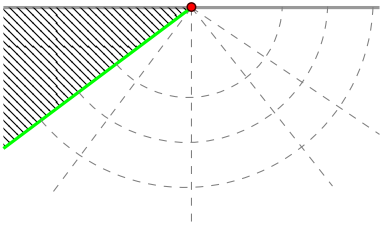
\begin{figure}
\begin{centering}
\begin{tikzpicture}[scale=1]
\draw[-,very thick,black!40] (-2.5,0) to (0,0);
\draw[-,very thick,black!40] (0,0) to (2.5,0);
\draw[pattern=north west lines,pattern color=black!200,draw=none] (0,0) to (-2.5,-1.875) to (-2.5,0) to (0,0);
\draw[-,dashed,color=black!50] (0,0) to (-1.875,-2.5); 
\draw[-,dashed,color=black!50] (0,0) to (0,-2.875);
\draw[-,dashed,color=black!50] (0,0) to (1.875,-2.375); 
\draw[-,dashed,color=black!50] (0,0) to (2.5,-1.6875); 
\draw[-,very thick,color=green!!50] (0,0) to (-2.5,-1.875);
\node at (0,0) {\textcolor{red}{$\bullet$}};
\node at (0,0) {\textcolor{black}{$\circ$}};
\draw[-,dashed,color=black!50] (-2,-1.5) arc (-140:-2:2.5);
\draw[-,dashed,color=black!50] (-1.5,-1.125) arc (-140:-2:1.875);
\draw[-,dashed,color=black!50] (-1,-0.75) arc (-140:-2:1.25);
\end{tikzpicture}
\caption{A constant time slice of an AdS$_{d+1}$ with a Karch-Randall brane. The green surface denotes the brane and it has AdS$_{d}$ geometry. The grey-shaded region behind the brane is cutoff.}
\label{pic:KR}
\end{centering}
\end{figure}

In the Karch-Randall (KR) braneworld, the graviton localized on the AdS$_{d}$ brane is identified with the lightest Kaluza-Klein (KK) mode of the ambient AdS$_{d+1}$ graviton (see Fig.~\ref{pic:KR}). It has been shown in the original work from Karch and Randall \cite{Karch:2000ct} that the mass of this mode is nonzero and it scales with the AdS$_{d}$ cosmological constant $\Lambda_{d}$ as $m_{0}^{2}\sim\Lambda_{d}^{2}$.  Porrati \cite{Porrati:2000cp} showed how this mass arises at one loop in the holographic dual theory. The holographic dual of the KR braneworld is a gravitational AdS$_{d}$ spacetime coupled to a nongravitational bath.  Ref.~\cite{Porrati:2001gx,Porrati:2002dt,Porrati:2003sa,Duff:2004wh,Aharony:2006hz,Geng:2023ynk,Geng:2023zhq} showed that in this scenario the AdS$_{d}$ graviton is massive due to spontaneous diffeomorphism breaking induced by the bath. Furthermore, it was shown by Porrati in \cite{Porrati:2000cp} that the above scaling of the mass with the d-dimensional cosmological constant leads to the absence of the van Dam-Veltman-Zakharov (vDVZ) discontinuity \cite{vanDam:1970vg,Zakharov:1970cc} for the AdS$_{d}$ graviton propagator in the zero graviton mass limit. The essence of the vDVZ discontinuity is that the massive graviton propagator doesn't reduce to the massless graviton propagator in the zero mass limit in flat space. This would be observable  since, for example,  the flat space gravitational potential could differ by 25\% from that of massless gravity in a four-dimensional spacetime. The change in gravity reflects additional polarization modes in massive gravity compared to its massless counterpart. 

The original argument for the absence of the vDVZ discontinuity from Porrati applies only at the classical level.   Ref.~\cite{Dilkes:2001av} argued that this discontinuity in AdS space will re-emerge at the one-loop level due to the additional polarizations of a massive graviton compared to a truly massless one.  Ref. \cite{Karch:2001jb} argued that this is not the case. We consider the origin of the discrepancy here.

  In fact, at the quantum level, there are two questions one can ask. One is whether the one-loop partition function has a smooth limit when the graviton mass goes to zero, essentially asking if the number of degrees of freedom is unchanged. 
  This would be true if there is a Higgs mechanism generating the graviton mass. The other is whether the loop-corrected graviton propagator reduces to that of the massless graviton in the zero mass limit of the massive graviton propagator.

In this letter, we show that in the KR braneworld the number of degrees of freedom is indeed continuous in the massless limit. This is consistent with the expectation that the limit of a massive graviton, which obtains its mass through spontaneous diffeomorphism breaking, should be continuous, with the limit being a massless graviton and a massive vector. Hence, this result supports the proposed Higgs mechanism in the holographic dual description by Ref.~\cite{Porrati:2001gx}. Moreover, the extra vector degrees of freedom in the graviton propagator decouple from matter in the massless limit, even at the quantum level. This follows from the argument in \cite{Karch:2001jb}, which we review and refine towards the end of the paper.  
For these reasons, physical processes should reduce to that of a theory with a massless graviton in this limit.

The goal of this letter is to provide an explicit proof of the continuity of the number of degrees of freedom in the massless limit of the KR braneworld. Along the way, we clarify several issues in \cite{Duff:1975ik,Christensen:1979iy} such as the treatment of the conformal factor. We will use the framework developed in \cite{Mazur:1989by} for the one-loop gravitational path integral in the Lorentzian signature. This result demonstrates the nature of the graviton mass in the KR braneworld as arising from spontaneous symmetry breaking and can be taken as a nontrivial check of the consistency with its holographic dual description. At the end, we comment on the implication of our result to the physics of entanglement islands and the difference between the massless limit of the KR scenario and the Randall-Sundrum II setup \cite{Randall:1999vf}.

\section{Covariant Locally Localized Gravity}
In this section we derive the fully covariant equations of motion for the induced graviton multiplet on the KR brane. This will enable us to carry out the one-loop path integral over this graviton multiplet which we will do in the Supplementary Material with the results shown in the next section. The original analysis in \cite{Karch:2000ct} relies on various gauge fixings at the classical level. This obstructed a calculation of the one-loop partition function and it was not so obvious how to restore to a fully covariant description. In fact, as we will discuss in the Supplementary Material the gauge choice made in \cite{Karch:2000ct} doesn't survive the massless limit of KR where there is a Higgs mechanism owing to the (d+1)-dimensional bulk.

Let's take the AdS$_{d+1}$ metric to be
\begin{equation}
    ds^{2}=g_{\mu\nu}dx^{\mu}dx^{\nu}=d\rho^{2}+e^{2A(\rho)} \bar{g}_{ab} dx^{a}dx^{b}\,,
\end{equation}
where $\rho\in(-\infty,\infty)$, $e^{2A(\rho)}=\cosh^{2}\rho$, $\bar{g}_{ab}$ is the AdS$_{d}$ metric and we have set the AdS length scale $l_{\text{AdS}}=1$. The KR brane sits at a constant-$\rho$ slice $\rho=\rho_{B}<0$ with $\rho_{B}$ determined by the brane tension \cite{Geng:2023qwm} and the brane cuts off the spacetime region $\rho\in(-\infty,\rho_{B})$. The induced graviton on the AdS$_{d}$ was identified as the lightest KK mode of the bulk AdS$_{d+1}$ graviton \cite{Karch:2000ct}. Therefore, our goal is to derive the equation of motion obeyed by these lowest KK modes.

The AdS$_{d+1}$ graviton obeys the linearized Einstein's field equations
\begin{equation}
\begin{split}
&\mathcal{E}^{(d+1)}_{\mu\nu\alpha\beta}h^{\alpha\beta}\equiv\frac{1}{2}g_{\mu\nu}\nabla_{\alpha}\nabla_{\beta}h^{\alpha\beta}-\frac{1}{2}g_{\mu\nu}\Box h+\frac{1}{2}\nabla_{\mu}\nabla_{\nu}h\\&-\frac{1}{2}\nabla_{\alpha}\nabla_{\mu}h_{\nu}^{\alpha}-\frac{1}{2}\nabla_{\alpha}\nabla_{\nu}h_{\mu}^{\alpha}+\frac{1}{2}\Box h_{\mu\nu}+\frac{d}{2}(g_{\mu\nu}h-2h_{\mu\nu})=0\,.\label{eq:linearizedEF}
    \end{split}
\end{equation}

We now want to consider what the $d$-dimensional description will be. As has been argued, the graviton decomposes into a KK tower of massive modes. 
According to bulk general covariance, these massive modes should preserve the induced d-dimensional diffeomorphism invariance. Thus, their mass terms should be of the St\"{u}ckelberg form with massive vector fields playing the role of the Goldstone bosons. We define the vector field $V_{\mu}(x,\rho)$, which is actually the gravitational Wilson line \cite{Donnelly:2015hta,Donnelly:2016rvo,Donnelly:2017jcd,Donnelly:2018nbv,Giddings:2018umg,Giddings:2019hjc,Geng:2025rov}, whose components are
\begin{equation}
    \begin{split}
   V_{\rho}(x,\rho)&=\frac{1}{2}\int_{\rho}^{\infty} du h_{\rho\rho}(x,u)\,, \\ V_{i}(x,\rho)&=e^{2A(\rho)}\int_{\rho}^{\infty} du e^{-2A(u)}h_{\rho i}(x,u)\\&+\frac{e^{2A(\rho)}}{2}\int _{\rho}^{\infty} due^{-2A(u)}\int_{u}^{\infty} du' \partial_{i}h_{\rho\rho}(x,u')\,.\label{eq:Vdef}
    \end{split}
\end{equation}
This above results can be obtained by solving 
\begin{equation}
h_{\rho \mu}(x, \rho)=-\nabla_\rho V_\mu(x, \rho)-\nabla_\mu V_\rho(x, \rho)\,.
\end{equation}

This vector field is defined to transform appropriately in the bulk theory in Equ.~(\ref{eq:Dtrans}) below. 
As explained in detail in the Supplementary Material, the above definition of $V_{i}(x,\rho)$ guarantees that its lowest KK mode survives the zero mass limit.

We further define
\begin{equation}
    \tilde{h}_{ij}(x,\rho)= h_{ij}(x,\rho)+2e^{2A(\rho)}A'(\rho)\bar{g}_{ij}(x)V_{\rho}(x,\rho)\,.\label{eq:newh}
\end{equation}
The above new fields transform under the bulk diffeomorphism as
\begin{equation}
\begin{split}
    V_{\mu}(x,\rho)&\rightarrow V_{\mu}(x,\rho)-\epsilon_{\mu}(x,\rho)\,,\\
    \tilde{h}_{ij}(x,\rho)&\rightarrow \tilde{h}_{ij}(x,\rho)+\bar{\nabla}_{i}\epsilon_{j}(x,\rho)+\bar{\nabla}_{j}\epsilon_{i}(x,\rho)\,,\label{eq:Dtrans}
    \end{split}
\end{equation}
where $\bar{\nabla}_{i}$ is the covariant derivative with respect to the $d$-dimensional metric $\bar{g}_{ab}$. 

The bulk Einstein's equations Equ.~(\ref{eq:linearizedEF}) are decomposed as three independent sets of equations
\begin{equation}
\begin{split}
  \bar{\mathcal{E}}^{(d)}_{ijmn}\tilde{h}^{mn}+&\frac{1}{2}e^{-(d-4)A(\rho)}\partial_{\rho}\Big\{e^{dA(\rho)}\partial_{\rho}e^{-2A(\rho)}\Big[(\tilde{h}_{ij}-\bar{g}_{ij}\tilde{h})\\&+(\bar{\nabla}_{i}V_{j}+\bar{\nabla}_{j}V_{i}-\bar{g}_{ij}2\bar{\nabla}_{i}V^{i})\Big]\Big\}=0\,,\label{eq:ij}
    \end{split}
\end{equation}
\begin{equation}
\begin{split}
\partial_{\rho}\Big[e^{-2A(\rho)}\big[\partial_{i}\tilde{h}-\bar{\nabla}^{j}\tilde{h}_{ij}-\bar{\nabla}_{j}&\Big(\bar{\nabla}_{i}V^{j}+\bar{\nabla}^{j}V_{i}\Big)\\&+2\bar{\nabla}_{i}\bar{\nabla}_{j}V^{j}\big]\Big]=0\,,\label{eq:irho}
    \end{split}
\end{equation}
and
\begin{equation}
    \frac{1}{2}e^{-(d-4)A(\rho)}\partial_{\rho}\Big[e^{dA(\rho)}\partial_{\rho}e^{-2A(\rho)}(\tilde{h}+2\bar{\nabla}_{i}V^{i})\Big]=0\,,\label{eq:rhorho}
\end{equation}
with all upper indices lifted from lower ones using $\bar{g}^{ij}$.

Now we can do the KK decomposition of the fields $V_{a}(x,\rho)$ and $\tilde{h}_{ab}(x,\rho)$ as
\begin{equation}
    V_{a}(x,\rho)=\sum_{n}\phi_{n}(\rho) V^{(n)}_{a}(x)\,,\quad \tilde{h}_{ab}(x,\rho)=\sum_{n}\phi_{n}(\rho)\tilde{h}^{(n)}_{ab}(x)\,,
\end{equation}
where the KK wavefunctions $\phi_{n}(\rho)$ satisfy the following eigenvalue equation and boundary condition
\begin{equation}
\begin{split}
    -e^{-(d-4)A(\rho)}\partial_{\rho}e^{dA(\rho)}\partial_{\rho}e^{-2A(\rho)} \phi_{n}(\rho)&=m_{n}^{2}\phi_{n}(\rho)\,,\\\partial_{\rho}e^{-2A(\rho)}\phi_{n}(\rho)|_{\rho=\rho_{B}}&=0\,,\label{eq:KKeomgraviton}
    \end{split}
\end{equation}
where $m_{n}^2$ is the mass square for the $n$-th KK mode.\footnote{We note that here for convenience we have considered the metric on AdS$_{d}$ to be $\bar{g}_{ab}$. In fact, in the KR braneworld one would like the AdS$_{d}$ geometry to be that on the KR brane which is $e^{2A(\rho_{B})}\bar{g}_{ab}$, where $\rho_{B}$ is the place of the KR brane in the $(d+1)$-dimensional bulk. The graviton mass in this metric of the brane is $e^{-2A(\rho_{B})}m_{n}^2$ and it is this mass that obeys $m_{n}^2\sim\Lambda_{d}^2$ for $n=0$ with $\Lambda_{d}$ the cosmological constant of the KR brane.} From the eigenvalue equation, we can derive the orthogonality condition of the wavefunctions $\phi_{n}(\rho)$
\begin{equation}
    \int_{\rho_{B}}^{\infty} d\rho e^{(d-6)A(\rho)}\phi_{n}(\rho)\phi_{m}(\rho)=\delta_{nm}\,,
\end{equation}
for normalized wavefunctions. \sout{As a result} Hence, one can see that the zero mass eigenmode $\tilde{\phi}_{0}(\rho)=e^{2A(\rho)}=\cosh^{2}\rho$ is not normalizable. As a result, the physical modes are all massive with nonzero mass squares $m_{n}^{2}>0$. We are here interested only in the lowest normalizable eigenmode whose wavefunction is denoted as $\phi_{0}(\rho)$ with eigenvalue $m_{0}^{2}$. So we have the equations of motion
\begin{equation}
\begin{split}
  \bar{\mathcal{E}}^{(d)}_{ijmn}\tilde{h}^{(0)mn}-\frac{1}{2}m_{0}^{2}&\Big[(\tilde{h}^{(0)}_{ij}-\bar{g}_{ij}\tilde{h}^{(0)})\\&+(\bar{\nabla}_{i}V^{(0)}_{j}+\bar{\nabla}_{j}V^{(0)}_{i}-\bar{g}_{ij}2\bar{\nabla}_{i}V^{(0)i})\Big]=0\,,\label{eq:ij0}
    \end{split}
\end{equation}
\begin{equation}
\begin{split}
\partial_{i}\tilde{h}^{(0)}-\bar{\nabla}^{j}\tilde{h}^{(0)}_{ij}-\bar{\nabla}_{j}\Big(\bar{\nabla}_{i}V^{(0)j}+\bar{\nabla}^{j}V^{(0)}_{i}\Big)+2\bar{\nabla}_{i}\bar{\nabla}_{j}V^{(0)j}=0\,,\label{eq:irho0}
    \end{split}
\end{equation}
and
\begin{equation}
\tilde{h}^{(0)}+2\bar{\nabla}_{i}V^{(0)i}=0\,,\label{eq:rhorho0}
\end{equation}
where Equ.~(\ref{eq:ij0}) and Equ.~(\ref{eq:irho0}) are exactly the equations of motion for an AdS$_{d}$ massive graviton $\tilde{h}^{(0)}_{ab}(x)$ with the St\"{u}ckelberg vector field $V^{(0)}_{a}(x)$ and Equ.~(\ref{eq:rhorho0}) is redundant with Equ.~(\ref{eq:ij0}) and Equ.~(\ref{eq:irho0}).\footnote{Equ.~(\ref{eq:irho0}) is obtained from Equ.~(\ref{eq:irho}) by applying $e^{-(d-4)A(\rho)}\partial_{\rho}e^{dA(\rho)}$ on the left-hand side and projecting the result to the lowest KK sector.} The above equations are invariant under the induced $d$-dimensional diffeomorphism transform:
\begin{equation}
\begin{split}
       V^{(0)}_{i}(x)&\rightarrow V^{(0)}_{i}(x)-\epsilon^{(0)}_{i}(x)\,,\\\tilde{h}^{(0)}_{ij}(x)&\rightarrow \tilde{h}^{(0)}_{ij}(x)+\bar{\nabla}_{i}\epsilon^{(0)}_{j}(x)+\bar{\nabla}_{j}\epsilon^{(0)}_{i}(x)\,,\label{eq:Dtransform0}
\end{split}
\end{equation}
where we have the transformation parameters 
\begin{equation}
    \epsilon^{(0)}_{a}(x)=\int_{\rho_{B}}^{\infty} d\rho e^{(d-6)A(\rho)}\phi_{0}(\rho)\epsilon_{a}(x,\rho)\,,
\end{equation}
as the lowest KK components of the bulk diffeomorphism transformation parameters $\epsilon_{\mu}(x,\rho)$. We will explain in the Supplementary Material that in the massless limit $m_{0}\rightarrow 0$ the bulk tells us that the correct normalization in $d$ dimensions is for the field $m_{0}V^{(0)}_{i}(x)$ to be fixed so that the kinetic term of the vector field remains in the action when we take the mass to zero. Furthermore, in the massless limit, this properly normalized field would be invariant under the diffeomorphism $\epsilon^{(0)}_{i}(x)$ and it cannot be gauged away as in the massive cases. 


\section{One-loop Partition Functions}
With the equations of motion at hand, we can write the effective action and compute the one-loop partition function. The details of the calculations are provided in the Supplementary Material and here we  provide only the results and their interpretations. The one-loop partition function is given by
\begin{equation}
    Z_{\text{massive}}=[\det\text{$^{\prime}$}(\Delta^{\text{TT}})]^{-\frac{1}{2}}\,,
\end{equation}
where $\Delta^{\text{TT}}$ means the operator $\Delta_{2}=\text{\ovA{$\Box$}}+2-m_{0}^{2}$ acting on a symmetric transverse traceless rank-two tensor field, and $\prime$ means projecting out the zero modes in the determinant. This result suggests that we have $\frac{d(d+1)}{2}-d-1=\frac{(d+1)(d-2)}{2}$ degrees of freedom which is the same as a massive graviton. Furthermore, a direct calculation of the one-loop partition in the massless limit gives the result
\begin{equation}
    Z_{\text{massless}}=[\det\text{$^{\prime}$}(\Delta^{\text{TT}})]^{-\frac{1}{2}}\,,
\end{equation}
where now $\Delta^{\text{TT}}$ means the operator $\Delta_{2}=\text{\ovA{$\Box$}}+2$, i.e. exactly the same as the zero mass extrapolation of the above massive results. Hence, in the massless limit we still have $\frac{(d+1)(d-2)}{2}$ propagating degrees of freedom which is the same as a massless graviton and a massive vector. Note that we are comparing two different calculations of the partition function itself, one for the cases that the graviton mass is nonzero and the other one for the theory in the massless limit. We are not simply taking the massless limit of the partition function result of the massive cases, for which the additional degrees of freedom necessarily survive the limit. This continuity distinguishes the massive gravity theory in the KR braneworld from the naive ones considered in \cite{Dilkes:2001av} and is consistent with the holographic dual interpretation that the graviton becomes massive due to the Higgs mechanism from the spontaneous breaking of the diffeomorphism. In essence, \cite{Dilkes:2001av} considered the gauge-fixed version of the our system with a gauge choice $V_{i}(x)=0$. As we explain in the Supplementary Material, this gauge choice is not consistent with the massless limit of KR, in which the the massive vector field $V^{(0)}_{i}(x)$ and the graviton $\tilde{h}^{(0)}_{ij}(x)$ decouple but the vector degrees of freedom survive.

\section{Massless Limit of the Loop-corrected Graviton Propagator}


We have not yet shown whether the decoupling of the extra vector degrees of freedom in the graviton propagator from the matter persists at the quantum level. 
We know that this decoupling happens at tree level when the graviton mass squared  goes to zero faster than the d-dimensional cosmological constant, which is indeed the case in the KR braneworld. However, \cite{Dilkes:2001av} questioned in their conclusion whether this would still hold at loop level, where all the parameters are renormalized.
In Ref. \cite{Karch:2001jb} it was argued, based on the extra-dimensional origin of the theory, that there cannot be a discontinuity in physical quantities when the mass of the graviton is taken to zero. In order to generate a nonzero deviation of the coupling from the massive case, we would need the induced d-dimensional theory to contain an operator that won't survive in the massless limit. However, such an additional operator cannot be generated at the loop level in the KR braneworld. This is because the structure of the induced d-dimensional theory, i.e. Equ.~(\ref{eq:ij}), Equ.~(\ref{eq:irho}), Equ.~(\ref{eq:rhorho}) and the fact that the (renormalized) graviton mass scales with the (renormalized) cosmological constant as $m_{0}^2\sim\Lambda_{d}^2$, is  sensitive only to the fact that in (d+1)-dimensions we have a diffeomorphism invariant gravitational theory in AdS. Hence, as long as the (d+1)-dimensional theory is still diffeomorphism invariant and lives in AdS after the quantum effects are incorporated, the structure of the induced d-dimensional theory and its implications, such as the massless limit decoupling of the extra vector degrees of freedom, would persist at the quantum level. This is because the diffeomorphism invariance guarantees the $(d+1)$-dimensional linearized Einstein's equation and so the KK analysis would be the same as before.

Additionally, one might worry that the massive vector degrees of freedom could be coupled to 
 the probe matter at higher orders in the effective action. These higher order terms could be generated radiatively by the graviton. However, because we have a nonsingular limit of the graviton propagator when $\Lambda$ and hence $m_{0}$ goes to zero, any higher dimension operator it radiatively generates can have only coefficients with positive powers of these quantities (presumably with negative powers of the Planck mass 
$M_{P}$ and so in the massless limit, the contributions from such operators would vanish.

\section{Conclusions}
In this letter we showed that there is no vDVZ discontinuity for the induced gravity in the KR braneworld at the quantum level. The massive theory reduces to the massless theory and a decoupled massive vector field.


In principle, these extra vector degrees of freedom could be detected cosmologically if they are sufficiently populated. This vector is coupled only gravitationally according to the equivalence principle and does not directly couple to other matter fields.  Detecting such degrees of freedom through their effect on the Universe's expansion  in the early Universe, for example,  could in principle distinguish this theory from the simple Randall-Sundrum type construction. 

Our derivation relies on a  fully covariant construction. The result allows us to compute the one-loop partition function and thus count the number of independent degrees of freedom in both the massive case and the massless limit. We found a continuous result for the one-loop partition in the massless limit and the result shows that the number of propagating degrees of freedom is conserved before and after we take the massless limit. This is consistent with the holographic dual description that the graviton mass is generated through a Higgs mechanism associated with the spontaneous diffeomorphism breaking and is a nontrivial check of the holographic duality.

 The KR braneworld plays an important role in the recent progress of the black hole information problem \cite{Almheiri:2019hni,Almheiri:2019psy,Geng:2020qvw,Chen:2020uac,Chen:2020hmv} as it allows for a simple constructions of the so-called entanglement island \cite{Penington:2019npb,Almheiri:2019psf} and the calculation of the unitary Page curve. Our result in this paper is consistent with the observation that the spontaneous diffeomorphism breaking is essential for the existence of islands. \cite{Geng:2020qvw,Geng:2020fxl,Geng:2021hlu,Geng:2025rov,Geng:2025rov,Geng:2025gqu}. Since diffeomorphism should be restored in the massless limit so we would expect islands to disappear.  

\begin{figure}[h]
\begin{centering}
\begin{tikzpicture}[scale=1.4]
\draw[-,very thick,black!100] (-2,0) to (0,0);
\draw[-,very thick,black!100] (0,0) to (1.25,0);
\draw[-,very thick,orange] (1.25,0) to (2.45,0);
\draw[pattern=north west lines,pattern color=purple!200,draw=none] (0,0) to (-2,-1.5) to (-2,0) to (0,0);
\draw[-,very thick,color=green!!50] (0,0) to (-2,-1.5);
\node at (0,0) {\textcolor{red}{$\bullet$}};
\node at (0,0) {\textcolor{black}{$\circ$}};
\draw[-,very thick,blue] (-1,-0.75) to (-2,-1.5);
\node at (-1,-0.75) {\textcolor{black}{$\bullet$}};
\node at (-0.9,-1) {\textcolor{black}{$\partial\mathcal{I}$}};
\draw[-,thick,color=red] (-1,-0.75) arc (-140:-1:1.25);
\draw[-,thick,color=red] (1.22,0) to (1.23,-1.5);
\node at (0,-1.49) {\textcolor{red}{$\gamma_{I}$}};
\node at (1.5,-0.75) {\textcolor{red}{$\gamma_{II}$}};
\end{tikzpicture}
\caption{\small A demonstration of the construction of an entanglement island in the KR braneworld. The orange line denotes the subregion for which we are looking to compute the entanglement entropy, with the blue region on the brane denoting its entanglement island $\mathcal{I}$. The red surface $\gamma_{I}$ connecting the asymptotic boundary and $\partial\mathcal{I}$ is a minimal area surface. $\gamma_{I}$ is determined by minimizing its area also with respect to its ending point $\partial \mathcal{I}$ on the brane. $\gamma_{I}$ and $\gamma_{II}$ are called the Ryu-Takayanagi surfaces.}
\label{pic: branwisland}
\end{centering}
\end{figure}
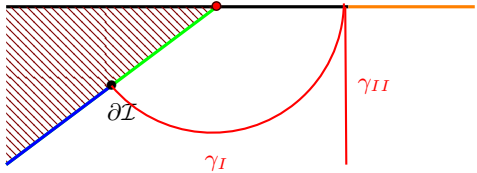

 In the KR braneworld (see Fig.~\ref{pic: branwisland}), the island can be constructed by finding the entanglement entropy of a subregion of the leftover asymptotic boundary of the AdS$_{d+1}$ bulk. This entropy can be computed using the Ryu-Takayanagi formula \cite{Ryu:2006bv}. In this scenario, there are two candidate types of RT surfaces due to the existence of the brane on which there could be replica wormholes \cite{Geng:2024xpj,Geng:2025efs}. In the general massive case, we have to compute the area of these two types of RT surfaces and take the lower area one to be the one that computes the entanglement entropy. The area of the two types of RT surfaces depicted in Fig.~\ref{pic: branwisland} are both UV divergent due to their starting point at the asymptotic boundary. The island emerges only if the RT surface that ends on the brane has a smaller area. However, in the small mass/small $d$-dimensional cosmological constant limit, the brane moves very close to the cutoff asymptotic boundary. That is the shaded-region in Fig.~\ref{pic:KR} is shrinking. Thus, the RT surface that ends on the brane,  the one that gives the island, would have an additional large contribution to its area compared with the one that doesn't end on the brane.  As a result, the islands necessarily shrink as we take smaller $d$-dimensional cosmological constant and disappear in the massless limit.  Here we see the island size is correlated with the degree to which diffeomorphism invariance is broken.

\section*{Acknowledgements}
We would like to thank Tom Hartman, Juan Maldacena, Andreas Karch, Massimo Porrati and Suvrat Raju for discussions. We thank Andreas Karch and Massimo Porrati for comments on a draft of this letter. HG and LR are supported by the Gravity, Spacetime, and Particle Physics (GRASP) Initiative from Harvard University. MM is supported by the Harvard Quantum Initiative. 

\bibliographystyle{apsrev4-1}
\bibliography{main}

\pagebreak

\widetext
\begin{center}
\textbf{\large Supplemental Material}
\end{center}

\setcounter{equation}{0}
\setcounter{figure}{0}
\setcounter{table}{0}
\setcounter{page}{1}
\makeatletter
\renewcommand{\theequation}{S\arabic{equation}}
\renewcommand{\thefigure}{S\arabic{figure}}
\renewcommand{\bibnumfmt}[1]{[S#1]}
\renewcommand{\citenumfont}[1]{S#1}
In this supplementary section, we address several subtleties that we pondered in the main text. 
\section{Details in the Computation of the One-loop Partition Function}
\subsection{Massive Case}
The equations of motion Equ.~(\ref{eq:ij0}) and Equ.~(\ref{eq:irho0}) tell us that the effective action of the induced graviton multiplet on the AdS$_{d}$ is
\begin{equation}
\begin{split}
    S_{\text{eff}}[\tilde{h}_{ab},V_{a}]=\frac{1}{2}\int d^{d}x\sqrt{-g}\Big[\tilde{h}\bar{\nabla}_{a}\bar{\nabla}_{b}\tilde{h}^{ab}-&\tilde{h}\text{\ovA{$\Box$}}\tilde{h}+\tilde{h}^{ab}\bar{\nabla}_{a}\bar{\nabla}_{b}\tilde{h}-2\tilde{h}^{ab}\bar{\nabla}_{c}\bar{\nabla}_{a}\tilde{h}_{b}^{c}+\tilde{h}^{ab}\text{\ovA{$\Box$}}\tilde{h}_{ab}+(d-1)(\tilde{h}^{2}-2\tilde{h}^{ab}\tilde{h}_{ab})\\&-m_{0}^{2}(\tilde{h}_{ab}+\bar{\nabla}_{a}V_{b}+\bar{\nabla}_{b}V_{a})(\tilde{h}^{ab}+\bar{\nabla}^{a}V^{b}+\bar{\nabla}^{b}V^{a})+m_{0}^{2}(\tilde{h}+2\bar{\nabla}^{c}V_{c})^{2}\Big]\,,\label{eq:action}
    \end{split}
\end{equation}
where for simplicity we have omitted the upper index $^{(0)}$ for $\tilde{h}_{ab}$ and $V_{a}$. To compute the one-loop partition we will evaluate the Lorentzian path integral
\begin{equation}
    Z=\int \frac{D[\tilde{h}_{ab}] D[V_{a}]}{\text{Vol(Diff)}} e^{iS_{\text{eff}}[\tilde{h}_{ab},V_{a}]}\,,\label{eq:partition}
\end{equation}
where the diffeomorphism gauge group acts as Equ.~(\ref{eq:Dtransform0}). We can simplify the path integral by the following redefinition
\begin{equation}
    \tilde{h}_{ab}\rightarrow\tilde{h}_{ab}+\bar{\nabla}_{a}V_{b}+\bar{\nabla}_{b}V_{a},\quad V_{a}\rightarrow V_{a}\,,
\end{equation}
which, by the diffeomorphism invariance of the linearized Einstein-Hilbert action, gives\footnote{An equivalent to get the same result is to do the Faddeev-Popov quantization with the gauge choice $V^{a}=0$.}
\begin{equation}
     Z=\int \frac{D[\tilde{h}_{ab}] D[V_{a}]}{\text{Vol(Diff)}} e^{iS_{\text{eff}}[\tilde{h}_{ab},0]}=\int D[\tilde{h}_{ab}] e^{iS_{\text{eff}}[\tilde{h}_{ab},0]}\,.\label{eq:partitionsimp}
\end{equation}
To compute this one-loop partition function, let's then do the York decomposition \cite{York:1973ia} of the metric $\tilde{h}_{ab}$
\begin{equation}
    \tilde{h}_{ab}=\frac{1}{d}\tilde{h}\bar{g}_{ab}+(L\xi)_{ab}+\tilde{h}^{\text{TT}}_{ab}\,,\label{eq:hrepara}
\end{equation}
where $\tilde{h}^{\text{TT}}_{ab}$ is the transverse traceless part and the longitudinal traceless part is given by
\begin{equation}
(L\xi)_{ab}=\bar{\nabla}_{a}\xi_{b}+\bar{\nabla}_{b}\xi_{a}-\frac{2}{d}\bar{g}_{ab}\bar{\nabla}^{c}\xi_{c}\,.
\end{equation}
Following \cite{Mazur:1989by}, let's pick the ultralocal measure
\begin{equation}
    \langle h,h\rangle=\int d^{d}x\sqrt{-g}G^{abcd}h_{ab}h_{cd}\,,\label{eq:ultra}
\end{equation}
where 
\begin{equation}
    G^{abcd}=\frac{1}{2}(\bar{g}^{ac}\bar{g}^{bd}+\bar{g}^{ad}\bar{g}^{bc}+C\bar{g}^{ab}\bar{g}^{cd})\,,\label{eq:G}
\end{equation}
with $C$ a constant to be discussed later. Under this measure, we can find the conjugate of the operator $L$ using
\begin{equation}
    \langle h,L\xi\rangle=\langle L^{\dagger}h,\xi\rangle\,,
\end{equation}
as
\begin{equation}
    (L^{\dagger}h)_{a}=-2\bar{\nabla}^{b}h_{ab}\,,
\end{equation}
where we note that $L^{\dagger}$ only acts on symmetric traceless rank two tensors.

As a result, the path integral measure under the decomposition becomes
\begin{equation}
    D[\tilde{h}_{ab}]=D[\tilde{h}^{\text{TT}}_{ab}]D[\tilde{h}]D[\xi_{a}][\det\text{$^{\prime}$}(L^{\dagger}L)]^{\frac{1}{2}}\,,\label{eq:measure}
\end{equation}
where in AdS$_{d}$ we have
\begin{equation}
    (L^{\dagger}L)_{a}^{b}=-2\Big[\text{\ovA{$\Box$}}\delta_{a}^{b}+\frac{d-2}{d}\bar{\nabla}_{a}\bar{\nabla}^{b}-(d-1)\delta_{a}^{b}\Big]\,.
\end{equation}
Moreover, under the decomposition Equ.~(\ref{eq:hrepara}) the effective action becomes
\begin{equation}
    \begin{split}
2S_{\text{eff}}[\tilde{h}_{ab},0]&=\langle\tilde{h}^{\text{TT}},\Delta_{2} \tilde{h}^{\text{TT}}\rangle-(1-\frac{3}{d}+\frac{2}{d^{2}})\langle \tilde{h},\text{\ovA{$\Box$}}\tilde{h}\rangle+(-\frac{2(d-1)}{d}+d-1+m_{0}^{2}\frac{d-1}{d})\langle \tilde{h},\tilde{h}\rangle\,,\\&+\langle \xi,L^{\dagger}\Delta_{2} L\xi\rangle+\frac{1}{2}\langle L^{\dagger}L\xi,L^{\dagger}L\xi\rangle+(1-\frac{2}{d})\langle \bar{\nabla}\tilde{h},L^{\dagger}L\xi\rangle,\label{eq:action2}
    \end{split}
\end{equation}
where we used the constraint Equ.~(\ref{eq:rhorho0}) and we defined
\begin{equation}
    \Delta_{2}\tilde{h}_{ab}=\text{\ovA{$\Box$}}\tilde{h}_{ab}+2\tilde{h}_{ab}-m_{0}^{2}h_{ab}\,.
\end{equation}
To proceed, we have to simplify the $\xi$-dependent pieces of the action Equ.~(\ref{eq:action2}) and this can be done by Hodge decomposition of $\xi_{a}$ \cite{Mazur:1989by}
\begin{equation}
    \xi_{a}=\partial_{a}\psi+\xi_{a}^{\text{H}}-\bar{\nabla}^{b}\omega_{ab}\,,\label{eq:decom}
\end{equation}
where $\psi$ is a scalar, $\xi^{\text{H}}_{a}$ is a harmonic one-form and $\omega_{ab}$ is a two-form so the last term is the transverse component which we will also denote as $\xi^{\text{T}}_{a}$. The following formulas will be useful for us
\begin{equation}
    \begin{split}
L^{\dagger}\Delta_{2}h_{ab}=\Delta'_{2}L^{\dagger}h_{ab}\,,\quad L^{\dagger}L\partial_{a}\psi=-4\frac{d-1}{d}\partial_{a}(\text{\ovA{$\Box$}}-d)\psi\,,\quad 
(L^{\dagger}L)\xi^{\text{T}}_{a}=\bar{\nabla}^{b}\Big(2(\text{\ovA{$\Box$}}-2)\Big)\omega_{ab}\,,\quad
\end{split}
\end{equation}
where $h_{ab}$ is any symmetric traceless tensor and $\Delta'_{2}=\text{\ovA{$\Box$}}-(d-1)-m_{0}^{2}$. For the above formulas we can simplify the action Equ.~(\ref{eq:action2}) as
\begin{equation}
\begin{split}
    2S_{\text{eff}}[\tilde{h}_{ab},0]=\langle\tilde{h}^{\text{TT}},\Delta_{2} \tilde{h}^{\text{TT}}\rangle-\frac{(d-1)(d-2)}{d^{2}}\langle (\tilde{h}-2\text{\ovA{$\Box$}}\psi),(\text{\ovA{$\Box$}}-d)(\tilde{h}-2\text{\ovA{$\Box$}}\psi)\rangle+m_{0}^{2}\frac{d-1}{d}\langle\tilde{h},\tilde{h}\rangle-m_{0}^{2}\langle \xi,L^{\dagger}L\xi\rangle\,,\label{eq:action3}
    \end{split}
\end{equation}
where we notice that in the above formula the harmonic one-form completely decouples which is a good news as in Equ.~(\ref{eq:measure}) we indeed don't have to worry about the kernel of $L^{\dagger}L$. Hereafter, we consider $\xi_{a}$ with only the scalar (longitudinal) component and the transverse component so we have
\begin{equation}
    \bar{\nabla}^{a}\xi_{a}=\text{\ovA{$\Box$}}\psi\,.
\end{equation}
To proceed, we use the measure in Equ.~(\ref{eq:measure}) in the path integral Equ.~(\ref{eq:partition}) by which the full path integral becomes
\begin{equation}
\begin{split}
    Z&=\int D[\tilde{h}^{\text{TT}}_{ab}]D[\tilde{h}]D[\xi_{a}][\det\text{$^{\prime}$}(L^{\dagger}L)]^{\frac{1}{2}} e^{iS_{\text{eff}}[\tilde{h}_{ab},0]}\,.\label{eq:partition2}
    \end{split}
\end{equation}
We can expand the last term in Equ.~(\ref{eq:action3}) as
\begin{equation}
    \langle L\xi,L\xi\rangle=\int d^{d}x\sqrt{-g} G^{abcd}(L\xi)_{ab}{L\xi}_{cd}=4\frac{d-1}{d}\langle\text{\ovA{$\Box$}}\psi,(\text{\ovA{$\Box$}}-d)\psi\rangle-2\langle\xi^{\text{T}},(\text{\ovA{$\Box$}}-(d-1))\xi^{\text{T}}\rangle\,,
\end{equation}
which gives the final form of the effective action
\begin{equation}
\begin{split}
2S_{\text{eff}}[\tilde{h}_{ab},0]=&\langle\tilde{h}^{\text{TT}},\Delta_{2} \tilde{h}^{\text{TT}}\rangle-\frac{(d-1)(d-2)}{d^{2}}\langle (\tilde{h}-2\text{\ovA{$\Box$}}\psi),(\text{\ovA{$\Box$}}-d)(\tilde{h}-2\text{\ovA{$\Box$}}\psi)\rangle+m_{0}^{2}\frac{d-1}{d}\langle\tilde{h},\tilde{h}\rangle\\&-4m_{0}^{2}\frac{d-1}{d}\langle\text{\ovA{$\Box$}}\psi,(\text{\ovA{$\Box$}}-d)\psi\rangle+2m_{0}^{2}\langle\xi^{\text{T}},(\text{\ovA{$\Box$}}-(d-1))\xi^{\text{T}}\rangle\,.\label{eq:actionfinal}
    \end{split}
\end{equation}
The last step before we are able to compute the one-loop partition function Equ.~(\ref{eq:partition2}) is to transform the path integration measure 
\begin{equation}
    D[\xi_{a}][\det\text{$^{\prime}$}(L^{\dagger}L)]^{\frac{1}{2}} =D[\xi_{a}][\det\text{$^{\prime}$}(\Delta^{\text{L}})\det\text{$^{\prime}$}(\Delta^{\text{T}})]^{\frac{1}{2}}=D[\psi]D[\xi^{\text{T}}_{a}][\det\text{\ovA{$\Box$}}]^{\frac{1}{2}}[\det\text{$^{\prime}$}(\Delta^{\text{L}})\det\text{$^{\prime}$}(\Delta^{\text{T}})]^{\frac{1}{2}}\,,\label{eq:reduce}
\end{equation}
where $\Delta^{\text{L}}=\text{\ovA{$\Box$}}-d$ is a scalar Laplacian operator and $\Delta^{\text{T}}=(\text{\ovA{$\Box$}}-(d-1))$ is a vector Laplacian operator which acts only on transverse vectors, i.e. $(d-1)$-components. 
As a result, the one-loop partition function is given by
\begin{equation}
\begin{split}
    Z=&[\det\text{\ovA{$\Box$}}]^{\frac{1}{2}}[\det\text{$^{\prime}$}(\Delta^{\text{L}})\det\text{$^{\prime}$}(\Delta^{\text{T}})]^{\frac{1}{2}}[\det\text{$^{\prime}$}(\Delta^{\text{TT}})\det\text{$^{\prime}$}(\Delta^{\text{T}})\det\text{\ovA{$\Box$}}\det\text{$^{\prime}$}(\Delta^{\text{L}})]^{-\frac{1}{2}}\,,\\=&[\det\text{$^{\prime}$}(\Delta^{\text{TT}})]^{-\frac{1}{2}}\,.\label{eq:resultmassive}
    \end{split}
\end{equation}
This result suggests that we have $\frac{d(d+1)}{2}-d-1=\frac{(d+1)(d-2)}{2}$ degrees of freedom which is the same as a massive graviton.


Moreover, we notice that the second term in Equ.~(\ref{eq:actionfinal}) has a different sign comparing to the other two kinetic terms so this suggests a conformal factor problem if we naively go to Euclidean signature as \cite{Gibbons:1978ac}. In fact, since we are following the Lorentzian approach, this issue is easily resolved by a proper choice of the measure, i.e. the constant C in Equ.~(\ref{eq:G}). We note that the measure Equ.~(\ref{eq:ultra}) defines the path integral measure $D[\tilde{h}]$ as
\begin{equation}
    \int D[\tilde{h}] e^{i\int d^{d}x\sqrt{-g}\frac{1}{2d}(2+Cd)\tilde{h}^{2}}=1\,,
\end{equation}
which is opposite to that of $\tilde{h}^{\text{TT}}_{ab}$
\begin{equation}
    \int D[\tilde{h}^{\text{TT}}_{ab}] e^{i\int d^{d}x\sqrt{-g}\tilde{h}^{\text{TT}}_{ab}\tilde{h}^{\text{TT}ab}}=1\,,
\end{equation}
if we take $C<-\frac{2}{d}$. Therefore, the conformal factor issue mentioned in \cite{Duff:1975ik,Christensen:1979iy,Dilkes:2001av} is resolved in our Lorentzian treatment by the choice $C<-\frac{2}{d}$ \cite{Mazur:1989by}.

\subsection{Massless Case}
Under the canonical normalization, the massless limit of the action Equ.~(\ref{eq:action}) is
\begin{equation}
    m_{0}^{2}\rightarrow0\,,\quad\text{$m_{0}V_{a}(x)$ fixed}\,,\label{eq:masslesslimit}
\end{equation}
which is consistent with the fact that the vector mode is the Goldstone mode from the holographic dual description. We will explain in the next section that this canonical normalization can be derived from the bulk. In this limit, $V_{a}(x)$ and $\tilde{h}_{ab}(x)$ are decoupled and the action becomes
\begin{equation}
\begin{split}
    S_{\text{eff}}[\tilde{h}_{ab},V_{a}]=\frac{1}{2}\int d^{d}x\sqrt{-g}\Big[\tilde{h}\bar{\nabla}_{a}\bar{\nabla}_{b}\tilde{h}^{ab}-&\tilde{h}\text{\ovA{$\Box$}}\tilde{h}+\tilde{h}^{ab}\bar{\nabla}_{a}\bar{\nabla}_{b}\tilde{h}-2\tilde{h}^{ab}\bar{\nabla}_{c}\bar{\nabla}_{a}\tilde{h}_{b}^{c}+\tilde{h}^{ab}\text{\ovA{$\Box$}}\tilde{h}_{ab}+(d-1)(\tilde{h}^{2}-2\tilde{h}^{ab}\tilde{h}_{ab})\\&-m_{0}^{2}(\bar{\nabla}_{a}V_{b}+\bar{\nabla}_{b}V_{a})(\bar{\nabla}^{a}V^{b}+\bar{\nabla}^{b}V^{a})+m_{0}^{2}(2\bar{\nabla}^{c}V_{c})^{2}\Big]\,,\label{eq:actionmassless}
    \end{split}
\end{equation} 
So the partition function reads
\begin{equation}
Z_{\text{massless}}=\int \frac{D[\tilde{h}_{ab}]D[V_{a}]}{\text{Vol(Diff)}}  e^{i S_{\text{eff}}[\tilde{h}_{ab},V_{a}]}\,,
\end{equation}
where the diffeomorphism now only acts on $\tilde{h}_{ab}$ because of the canonical normalization Equ.~(\ref{eq:masslesslimit}). We can do the same York decomposition for the metric and the Hodge decomposition as before in the massive case which gives
\begin{equation}
\begin{split}
       2S_{\text{eff}}[\tilde{h}_{ab},V_{a}]=&\langle\tilde{h}^{\text{TT}},\Delta_{2} \tilde{h}^{\text{TT}}\rangle-\frac{(d-1)(d-2)}{d^{2}}\langle (\tilde{h}-2\text{\ovA{$\Box$}}\psi),(\text{\ovA{$\Box$}}-d)(\tilde{h}-2\text{\ovA{$\Box$}}\psi)\rangle\\&-m_{0}^{2}\langle \bar{\nabla}_{a}V_{b}+\bar{\nabla}_{b}V_{a},\bar{\nabla}_{c}V_{d}+\bar{\nabla}_{d}V_{c}\rangle+(4-\frac{4}{d})m_{0}^{2}\langle \bar{\nabla}_{c}V^{c},\bar{\nabla}_{d}V^{d}\rangle\,.\label{eq:actionmassless3}
    \end{split}
\end{equation}
One can proceed with doing the Hodge decomposition also for the vector field $V^{a}(x)$. From the action Equ.~(\ref{eq:action}) we can see that the harmonic one-form part of $V^{a}(x)$ automatically decouples,\footnote{This is easy to see by noticing that the harmonic one-form is divergenceless.} so we have
\begin{equation}
    V_{a}(x)=\partial_{a}\psi_{V}+V^{\text{T}}_{a}(x)\,,
\end{equation}
with $V^{\text{T}}_{a}(x)$ denoting the transverse components. As a result, we have
\begin{equation}
\begin{split}
     2S_{\text{eff}}[\tilde{h}_{ab},V_{a}]=\langle\tilde{h}^{\text{TT}},\Delta_{2} \tilde{h}^{\text{TT}}\rangle-\frac{(d-1)(d-2)}{d^{2}}\langle (\tilde{h}-2\text{\ovA{$\Box$}}\psi),(\text{\ovA{$\Box$}}-d)(\tilde{h}-2\text{\ovA{$\Box$}}\psi)\rangle&+4m_{0}^{2}(d-1)\langle\psi_{V},\text{\ovA{$\Box$}}\psi_{V}\rangle\\&+2m_{0}^{2}\langle V^{\text{T}},(\text{\ovA{$\Box$}}-(d-1))V^{\text{T}}\rangle\,\,,\label{eq:actionmassless4}
     \end{split}
\end{equation}
where $\Delta_{2}h_{ab}=\text{\ovA{$\Box$}}\tilde{h}_{ab}+2\tilde{h}_{ab}$. The path integral measure for the metric can be decomposed as before
\begin{equation}
    D[\tilde{h}_{ab}]=D[\tilde{h}^{\text{TT}}_{ab}]D[\tilde{h}]D[\xi_{a}][\det\text{$^{\prime}$}(L^{\dagger}L)]^{\frac{1}{2}}=D[\tilde{h}^{\text{TT}}_{ab}]D[\tilde{h}]D[\xi_{a}][\det\text{$^{\prime}$}(\Delta^{\text{L}})\det\text{$^{\prime}$}(\Delta^{\text{T}})]^{\frac{1}{2}}\,,
\end{equation}
and the measure for the vector field $V^{a}$ can be decomposed as
\begin{equation}
    D[V^{a}]=D[\psi_{V}][\det\text{\ovA{$\Box$}}]^{\frac{1}{2}}D[V^{\text{T}}_{a}]\,.
\end{equation}
Now we can compute the partition function Equ.~(\ref{eq:partition2}) as
\begin{equation}
    Z_{\text{massless}}=\int D[\tilde{h}^{\text{TT}}_{ab}]D[\tilde{h}]\frac{D[\xi_{a}]}{\text{Vol(Diff)}}D[V^{\text{T}}_{a}]D[\psi_{V}][\det\text{\ovA{$\Box$}}]^{\frac{1}{2}}[\det\text{$^{\prime}$}(\Delta^{\text{L}})\det\text{$^{\prime}$}(\Delta^{\text{T}})]^{\frac{1}{2}}e^{iS_{\text{eff}}[\tilde{h}_{ab},V_{a}]}\,,
\end{equation}
where we can see that if we redefine the variable, which contributes a trivial Jacobian, $\tilde{h}\rightarrow\tilde{h}-2\text{\ovA{$\Box$}}\psi$ the longitudinal metric mode $\xi_{a}$ completely decouples in the path integral and so the path integral over $\xi_{a}$ cancels against the volume of the diffeomorphism group. As a result, we have
\begin{equation}
    \begin{split}
    Z=&[\det\text{\ovA{$\Box$}}]^{\frac{1}{2}}[\det\text{$^{\prime}$}(\Delta^{\text{L}})\det\text{$^{\prime}$}(\Delta^{\text{T}})]^{\frac{1}{2}}[\det\text{$^{\prime}$}(\Delta^{\text{TT}})\det\text{$^{\prime}$}(\Delta^{\text{T}})\det\text{$^{\prime}$}(\Delta^{\text{L}})\det\text{\ovA{$\Box$}}]^{-\frac{1}{2}}\,,\\=&[\det\text{$^{\prime}$}(\Delta^{\text{TT}})]^{-\frac{1}{2}}\,,
    \end{split}
\end{equation}
which is exactly the zero mass limit of Equ.~(\ref{eq:resultmassive}) with $\frac{d(d+1)}{2}-d-1=\frac{(d+1)(d-2)}{2}$ independently propagating modes. 

\subsection{A Subtlety of the Massless Limit}

We note that one potentially asks the question that what happens to the $\rho\rho$-component equation of motion of the (d+1)-dimensional bulk graviton. In the massive case, it tells us that
\begin{equation}
    m_{0}^2 (\tilde{h}+2\bar{\nabla}_{a}V^{a})=0\,.
\end{equation}
Thus, in the massless limit Equ.~(\ref{eq:masslesslimit}) this equation becomes trivial and so our treatment of the massless case above is consistent.

However, one might be confused and asks what happens to the $i\rho$-component of the bulk graviton equation of motion. The same consideration as the the above $\rho\rho$-component might suggest that the Goldstone vector field $V^{a}$ completely decouples. However, we note that in fact the $i\rho$-components of the bulk graviton equation of motion only involve one $\rho$-derivative \cite{Geng:2025rov} so its scaling with the graviton mass $m_{0}$ is really linear. As a result, the Goldstone boson survives the massless limit Equ.~(\ref{eq:masslesslimit}). 

More precisely, from Equ.~(\ref{eq:irho}) the $i\rho$-component of the bulk equation of motion is actually an equation of $\partial_{\rho}e^{-2A(\rho)}\tilde{h}_{ij}(x,\rho)$ and $\partial_{\rho}e^{-2A(\rho)}V_{i}(x,\rho)$. Thus, using Equ.~(\ref{eq:Vdef}) we can see that it is an equation of $\partial_{\rho}e^{-2A(\rho)}\tilde{h}_{ij}(x,\rho)$ and $e^{-2A(\rho)}h_{\rho i}(x,\rho)$ and $e^{-2A(\rho)}h_{\rho\rho}(x,\rho)$. Now projecting this equation to the lowest KK modes and using the fact that $\phi_{0}(\rho)$ is localized to the KR brane in the massless limit and the boundary condition in Equ.~(\ref{eq:KKeomgraviton}), we can see that the lowest KK graviton $\tilde{h}^{(0)}_{ij}(x)$ decouples from Equ.~(\ref{eq:irho}). But the vector field $V_{i}^{(0)}(x)$ with a proper normalization should stay dynamical. This is because as we just explained that the $V_{i}(x)$-dependent parts of Equ.~(\ref{eq:irho}) can be written using $e^{-2A(\rho)}h_{\rho i}(x,\rho)$ and $e^{-2A(\rho)}h_{\rho\rho}(x,\rho)$ with no $\rho$-derivatives acting on them. Therefore, none of the KK modes of that particular combination of $e^{-2A(\rho)}h_{\rho i}(x,\rho)$ and $e^{-2A(\rho)}h_{\rho\rho}(x,\rho)$ whose line integral gives $V_{i}(x,\rho)$ become decoupled. Thus, this bulk consideration tells us that the correct normalization in d-dimension should be the one ensuring that the vector modes are not decoupled in the massless limit, which is the massless limit we listed in Equ.~(\ref{eq:masslesslimit}). Furthermore, it also explains why the gauge choice in \cite{Karch:2000ct} doesn't survive the massless limit. This is because we have the bulk gauge transform
\begin{equation}
\begin{split}
    e^{-2A(\rho)}h_{\rho i}(x,\rho)&\rightarrow e^{-2A(\rho)}h_{\rho i}(x,\rho)+e^{-2A(\rho)}\nabla_{i}\epsilon_{\rho}(x,\rho)+e^{-2A(\rho)}\nabla_{\rho}\epsilon_{i}(x,\rho)\,,\\&= e^{-2A(\rho)}h_{\rho i}(x,\rho)+\partial_{\rho}e^{-2A(\rho)}\epsilon_{i}(x,\rho)+e^{-2A(\rho)}\partial_{i}\epsilon_{\rho}(x,\rho)\,.
    \end{split}
\end{equation}
Hence, using the same argument as above we can see that the zeroth KK mode of the bulk gauge transform $\epsilon_{i}(x,\rho)$ will not transform $e^{-2A(\rho)}
h_{\rho i}(x,\rho)$ in the massless limit so it doesn't transform any d-dimensional fields other than the lowest graviton $\tilde{h}^{(0)}_{ij}(x)$ either. As a result, one cannot use it to gauge away $V^{(0)}_{i}(x)$. We note that this is also consistent with the normalization we listed in Equ.~(\ref{eq:masslesslimit}) as the canonical field $m_{0}V_{a}^{(0)}(x)$ doesn't transform under the diffeomorphism so it cannot be gauged away by a diffeomorphism fixing condition. As we discussed in the main text, the fact justifies the proposed Higgs mechanism in the holographic dual description of the KR braneworld. In essence, the above discussion reveals that the general covariance of the (d+1)-dimensional bulk indicates the existence of such a Higgs mechanism. 

Moreover, the above consideration helps to distinguish the massless limit in the KR setup from the massless sector in the Randall-Sundrum II setup. In the later case, there is no additional vector degrees of freedom in the d-dimensional graviton multiplet but in the former case the vector part survives. This can be seen from the fact that the disappearance of the vector part in the massless sector would be due to the exact zero mass KK wavefunction $\tilde{\phi}_{0}(\rho)=e^{2A(\rho)}$ \cite{Geng:2025rov}, which however is different from the wavefunction of the lowest massive mode in the KR case even in the massless limit. This is because the zero mass KK wavefunction $\tilde{\phi}_{0}(\rho)=e^{2A(\rho)}$ is not normalizable and doesn't die off in the part of the asymptotic boundary which is not cutoff by the KR brane but the KK wavefunction of the massive modes should all go to zero in that part even in the massless limit. More explicitly, the absence of the exact zero mass mode in the Goldstone vector field 
can be seen as follows. From Equ.~(\ref{eq:Vdef}), we can see that the vector $V_{i}(x,\rho)$ has no KK component with wavefunction $e^{2A(\rho)}$. This is because this component requires that the result of the integrals in Equ.~(\ref{eq:Vdef}) be $\rho$-independent, which implies that the integrands should be zero, and so the resulting integrals are also zero.
\end{document}